\documentclass{edp-conf}
\usepackage{graphicx}

\DeclareRobustCommand{\element}{\relax\ifmmode\@tempswafalse
\else\@tempswatrue\fi\clearelargs\def\?{\phantom{0}}\@lement}
\def\@lement#1{\if#1[\expandafter\f@@dargs\else\druck@lement{#1}\fi}

\DeclareMathAlphabet{\mathsc}{OT1}{cmr}{m}{sc}
\def\testbx{bx}%
\DeclareRobustCommand{\ion}[2]{%
\relax\ifmmode
\ifx\testbx\f@series
{\mathbf{#1\,\mathsc{#2}}}\else
{\mathrm{#1\,\mathsc{#2}}}\fi
\else\textup{#1\,{\mdseries\textsc{#2}}}%
\fi}

\begin{document}

\TitreGlobal{AGN in their Cosmic Environment}

\title{X-ray emitting-absorbing media in Seyfert 1 galaxies} 
\author{ M. Mouchet }\address{DAEC, Observatoire de Paris-Meudon F-92190 Meudon ;
\email{martine.mouchet@obspm.fr}}
\secondaddress{University Denis-Diderot, F-75005 Paris}
\author{ A. Abrassart }\sameaddress{1}
\author{ D. Porquet }\sameaddress{1}
\secondaddress{DSM/DAPNIA/SAp, CE Saclay, F-91191 Gif sur Yvette}
\author{ A.-M. Dumont }\sameaddress{1}
\author{ S. Collin}\sameaddress{1}
\runningtitle{X-ray emitting-absorbing media in Seyfert 1}
\maketitle
\begin{abstract} 
Seyfert galaxies have been shown to exhibit a large variety of features
in their X-ray spectra from which the environment of the central engine 
can be deduced. We focus on the two following aspects: the Warm 
Absorber, mainly responsible of the soft X-ray properties, and the 
reprocessing/reflecting plasma medium at the origin of the iron K$\alpha$ 
fluorescent line.

The physical parameters and  the location of the so-called 
Warm Absorber (WA), a photoionized medium along the line of sight 
to the nuclear region, are more strongly constrained by optical coronal lines 
than by the oxygen edges observed in the soft X-rays and produced by the
WA. The photoionization models also predict the intensities of the
X-ray emission lines which are going to be detected with the new generation 
of X-ray satellites.

An alternative model to the relativistic accretion disc is proposed to
explain the profile of the X-ray iron K$\alpha$ line observed in the
Seyfert 1. This line can be formed in the framework of a 
quasi-spherical accretion of  optically thick clouds. An  optically thick 
photoionization code coupled with a Monte-Carlo code has been developed 
to compute the  entire spectrum from the IR to the hard X-rays for 
a close geometry with a large covering factor.  The multiple Compton 
reflections allow to reproduce the redshifted broad iron line as
detected in several Seyfert 1 galaxies and in MCG-6-30-15.  
 
\end{abstract}
\section{Main X-ray properties of AGN}

All types of AGN are bright X-ray emitters and a large number has been
discovered after their optical identification with an X-ray source.
A large fraction of their high luminosity, due to the release of 
gravitational energy in the central regions, is radiated in the X-rays
($\sim$ 20 to 30 \% for Seyfert 1 and 10\% for quasars (see  
review by Mushotzky 
et al. 1993)).   In this paper, X-ray properties of Seyfert 1 only 
are discussed.
The shape of the X-ray spectrum above 2\,keV is schematically 
described by a power law
(photon index of -1.7) with a high energy cut-off at 100 keV, and   
below 1\,keV it is 
absorbed by a warm medium. It also exhibits a hump at
20~keV and a soft X-ray excess, particularly strong in narrow line Seyfert 1 
galaxies.
A broad emission iron line at 6.4\,keV is also often seen (Fabian et al. 
1995).

The X-ray emission is rapidly variable: variations on a timescale of a 
few hours are often detected in Seyfert 1 and down to  100~s in MCG-6-30-15 
(e.g. Yaqoob et al. 1997).
Such short 
timescales indicate emitting regions smaller than 10$^{13}$\,cm. The detection 
of a 
strong correlation between the UV and the X-rays with a small or even null 
lag might favor an origin of the UV due to the reprocessing of the X-rays
from a source close to the black hole. However this simple scenario is 
in conflict with recent observations of the two 
sources NGC 7469 (Nandra et al. 1998) and NGC 5548
(Chiang et al. 1999). 
On longer timescales several objects including some Narrow Line Seyfert 1 
galaxies have exhibited two brightness states with different spectral shape, 
similarly to galactic black hole candidates (f.i. 1H0419-577, Guainazzi et al.
1998). Several phenomenological models
have been proposed to account for the bulk of the X-ray emission such
as the disk-corona model, the standard irradiated model, the dense blob model
and the dilute cloud model (see review by Collin \& Dumont 1996).

\section{The Warm Absorber}
\subsection{Spectral characteristics}
About half of the Seyfert 1 show absorption edges at $\sim$ 0.8\,keV
identified either to \ion{O}{vii} (0.74\,keV) and/or \ion{O}{viii} 
(0.87\,keV) (Reynolds
1997).
In addition a few emission lines have been detected in the X-rays 
(\ion{O}{vii}, \ion{O}{viii}, Fe-L, \ion{Ne}{ix}) and in the UV 
(\ion{Ne}{viii}, \ion{O}{vi}).
These features reveal the presence of  partially or totally photoionized 
medium located in the central regions, the so-called Warm Absorber (WA).

\subsection{Constraints brought by coronal lines}
With the aim of determining the physical parameters of the WA and its location,
we have constrained the range of possible values of these parameters by 
confronting the predictions of models with observational data 
(Porquet et al. 1999).
The absorption edges and the resonance lines beeing almost insensitive
to the density in the range 10$^{4}$\,-\,10$^{12}\,$cm$^{-3}$, we have used 
optical coronal lines  
([{\ion{Fe}{X}], [{\ion{Fe}{XIV}}]) to constrain this parameter.\\

\begin{figure}[h]
\includegraphics[width=5.7cm,angle=0]{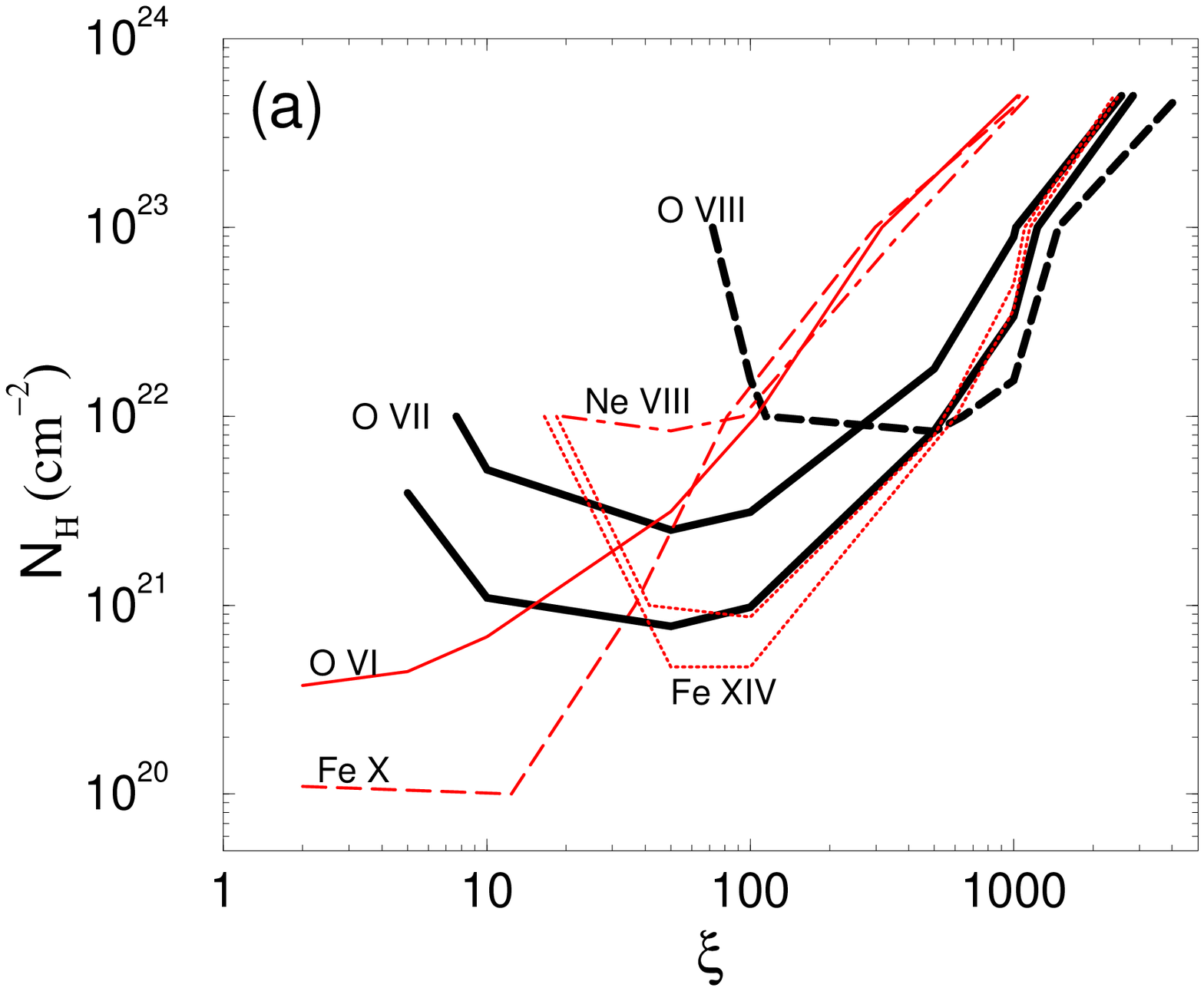}
\qquad
\includegraphics[width=5.7cm,angle=0]{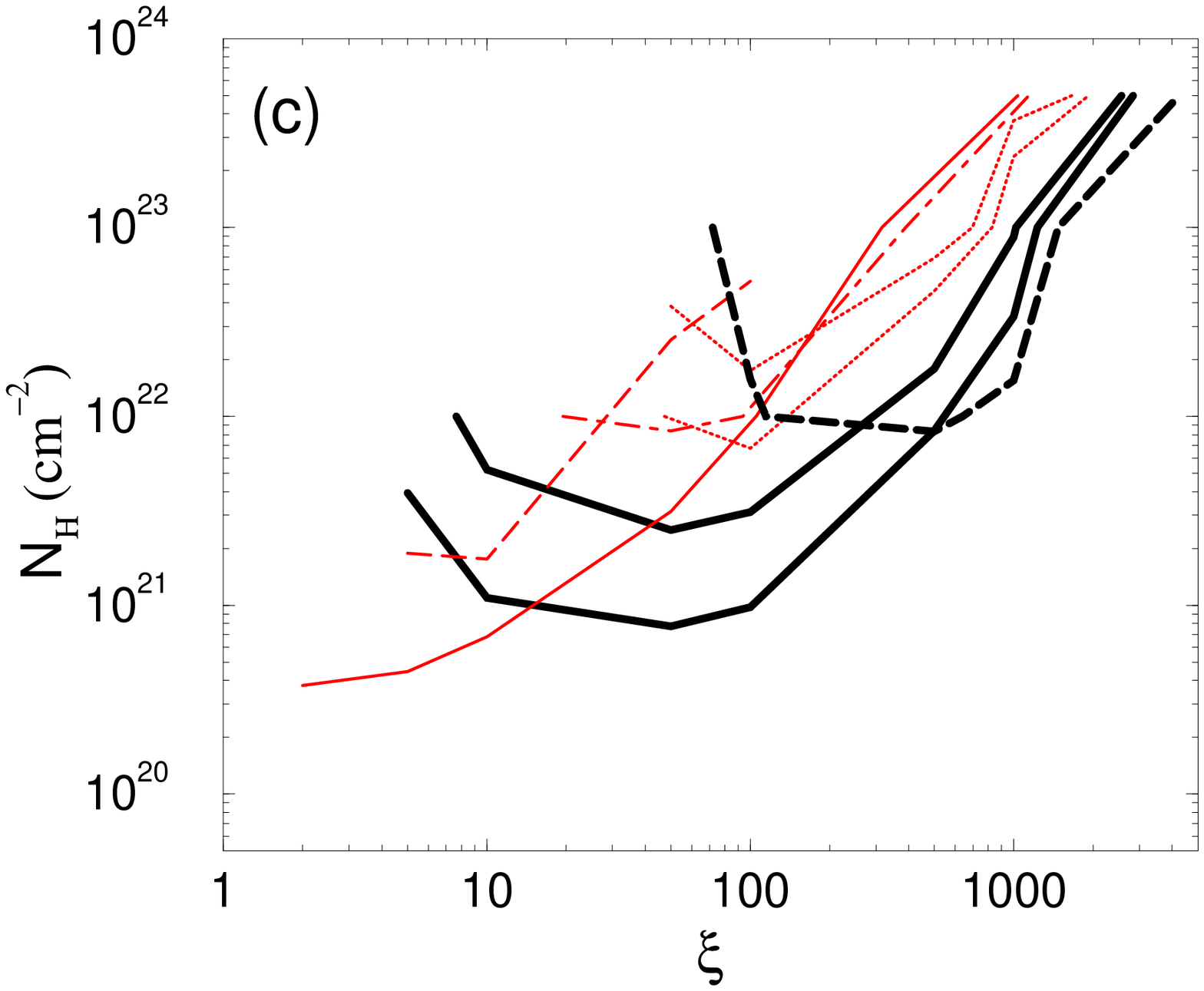}
\caption{Isovalue curves in the plane ($\xi$,N$_{\rm H}$) for the pure 
photoionisation model with the incident Laor continuum and for 
two densities:  n$_{\mathrm{H}}$=10$^{8}$\,cm$^{-3}$   (left) and 
n$_{\mathrm{H}}$=10$^{10}$\,cm$^{-3}$ (right). 
Thick lower and upper lines:
$<\tau_{O\,VII}>$=0.10 and 0.33 ; thick long dashed line:  $<\tau_{O\,VIII}>$=
0.20; thin long dashed line: EW([FeX])~=~1.5~\AA ; thin lower and upper dotted
lines: EW([FeXIV])~=~2~and~3\AA  ; thin dotted-dashed line: 
EW(Ne VIII)=4\AA{} and thin solid line: EW(OVI)=7\AA. 
 }
\end{figure}
Pure 
photoionisation models and hybrid models (radiative  plus collisional 
processes with fixed temperature) have been investigated for a large grid
of parameters. Model computation is done  by coupling two codes, one computing
the thermal and ionisation structure of photoionised optically thin clouds 
(PEGAS), the other code (IRIS) computing detailed multi-wavelength spectra 
of these clouds (Dumont \& Porquet 1998). The incident 
spectrum is that 
described by Laor et al. (1997). The ionisation state of the WA depends 
of the ionisation parameter 
$\xi~=~\frac{\mathrm{L}}{n_{\mathrm{H}}~\mathrm{R}^{2}}$  where L is the
 bolometric luminosity (erg\,s$^{-1}$),  
R the distance from the illuminated side of the cloud to the 
central source and n$_{\mathrm{H}}$ the density, assumed to be constant.\\
A covering factor of 0.5 was chosen, 
consistently with the proportion of Seyfert 1 showing soft X-ray absorption 
edges.  Input parameters are the column density  N$_{\mathrm{H}}$, the
density  n$_{\mathrm{H}}$, the $\xi$ parameter and the ionisation 
process/temperature.
Figure 1 is an illustration of the constraints brought by
the observational features on the above physical parameters. Average  
values of the optical 
depths and the equivalent widths are derived from the literature. 
Models giving rise to coronal lines stronger than observed are 
ruled out.
This confrontation of the predictions with the typical WA features and coronal
lines  observed in Seyfert 1 and also with those of the peculiar object 
MCG-06-30-15
leads to the two following main conclusions (Porquet et al. 1999).  First
the density of the warm absorber should be $> 10^{10}$\,cm$^{-3}$. Second, a 
two-zone
model is favoured, with an inner region located at a similar distance 
as the BLR ($\sim 10^{16}$\,cm) and responsible of the \ion{O}{viii} edge,
and an outer one associated with the \ion{O}{vii} edge, in agreement
with results found by  Otani et al. (1996) for MCG-6-30-15 based on 
variability studies.

\subsection{Diagnostics based on soft X-ray lines}
\begin{figure}[h]
\includegraphics[width=5.7cm,angle=0]{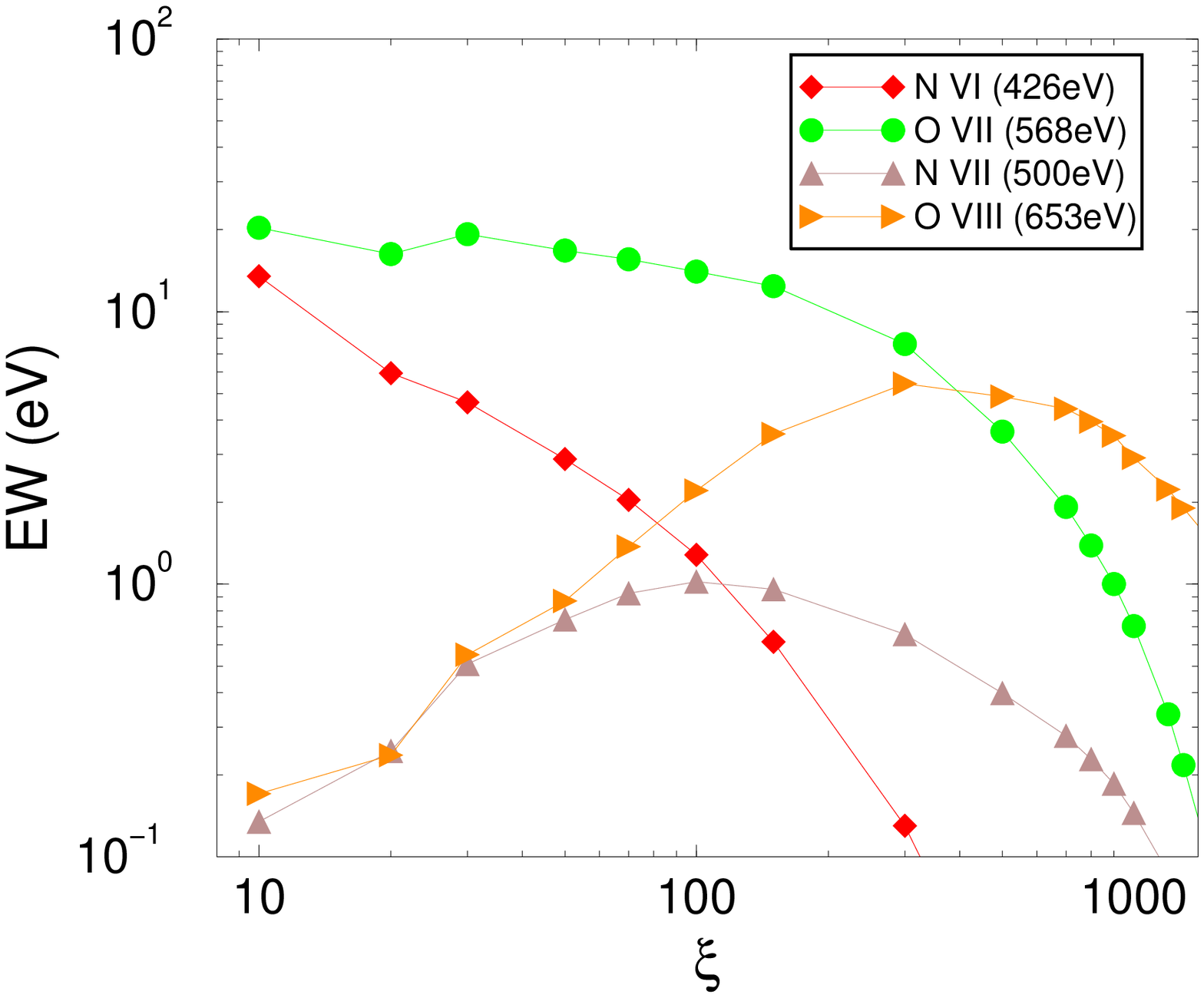}
\qquad
\includegraphics[width=5.7cm,angle=0]{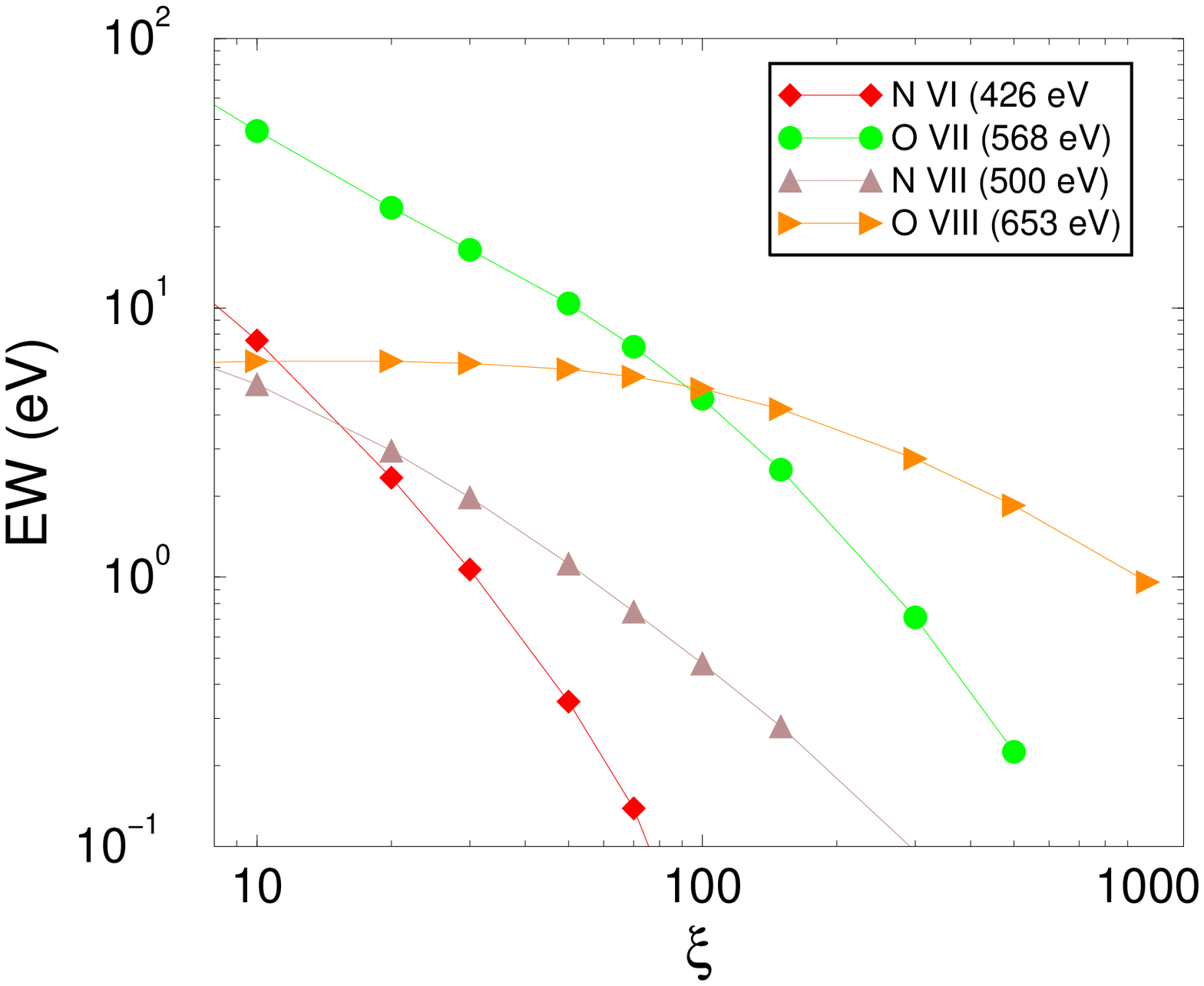}
\caption{Equivalent widths of soft X-ray emission resonance lines 
assuming an incident Laor continuum and a covering factor of 0.5,
with N$_{\rm H}$=10$^{22}$ cm$^{-2}$ and n$_{\rm H}$=10$^{10}$ cm$^{-3}$.
{\it Left}: pure photoionisation model ; {\it right}: hybrid model. } 
\end{figure}
Additional information will be soon available from the new generation of 
X-ray satellites (Chandra, XMM, Astro-E) which will allow the detection 
of emission lines thanks to their unprecedented high spectral resolution.

Figure 2 illustrates the strong influence of the ionisation parameter on the
equivalent widths (EW) of the main soft X-ray resonance lines. The EW  
have been computed both in the case of a pure photoionisation model and 
in a hybrid case for a fixed temperature of 10$^6$\,K.  In the first case,
the \ion{O}{vii} line is the strongest line over a wide range of $\xi$ 
parameter values while the  \ion{O}{viii} line dominates at high  $\xi$ 
($>$\,400 erg.cm.s$_1$). For the
hybrid model,   the \ion{O}{vii} line becomes dominant for a lower $\xi$ 
value ($>$\,100) (Porquet et al. 1998). However it should be noted that 
these results strongly depend on the turbulence velocity and on the 
functional form of the escape probability chosen to compute the line 
intensity, which depends on the fact that the redistribution in the line is 
partial or not.

\section{The Iron K$\alpha$ fluorescent line}
\subsection{Main characteristics}
A broad Fe line has been detected in 14 among 18 Seyfert 1 galaxies observed with ASCA
(Nandra et al. 1997), peaking at 6.4\,keV ( $\sim$ neutral iron), with
a typical
equivalent width of 150\,~eV and showing an asymmetric profile 
with an extended red wing.  The line intensity decreases for high X-ray 
luminosity. 

\subsection{Modeling}
To account for the Compton hump at $\sim$ 20\,keV, all models require 
the presence of a reflecting cold or mildly ionized medium illuminated by a 
hard 
X-ray continuum (typically with an energy power law index of -1), 
most probably originating
from thermal Compton up-scattering of soft seed photons.
The iron line is produced by fluorescence following the photoelectric 
absorption at the Fe K edge energy.
Two main explanations have been proposed to explain the red wing : either
the effects of gravitational and Doppler shifts or 
multi-Compton down-scattering.
The relativistic diskline model 
(Fabian et al. 1989) has been applied to different individual objects 
(Nandra et al. 1997, Iwasawa et al. 1996).
It consists of a thin disk illuminated by a hot corona or a central point
source. It can account for the Fe line profile in a Schwarzschild geometry 
or in a Kerr geometry for specific observations (such as during a flare
observed in MCG-6-30-15 (Iwasawa et al. 1999)), but provided that
the emissivity is  defined by a power law function of radius 
R$^{- \alpha}$,
$\alpha$ being a free parameter, and assuming a line at 6.4\,keV. 
An inclination 
angle of
 $\sim 30^{\circ}$ and a reflection factor close to 1 are requested. 
However this model relies on our knowledge of the very inner parts of 
the disc which is still uncertain as far as instabilities probably alter the 
density and ionisation structure.
     
\subsection{Multi Compton reflection in a closed geometry}
The Compton broadening has been first suggested as an alternative model 
 by Czerny et al. (1991)
to explain the Fe line in NGC 3227.
However cold material only produces a narrow red shoulder. A broad 
red wing requires repeated scattering, which are expected  to occur in
an ionised medium (Fabian et al. 1995). Broadening of an intrinsic 
narrow Fe 
line, transmitted through
a highly ionized Thompson thick cloud, has been proposed by Misra and 
Kembhavi (1998). This model has been ruled out by Reynolds \& Wilms (1999)
for  MCG-6-30-15 from the continuum variability timescale and the absence 
of an observed soft excess.
Here we describe a
different approach for which multiple scatterings arise in 
a closed shell geometry, ensuring  an amplification of the radiation inside 
the cloud 
system and thus the presence of highly photoionised clouds (Abrassart, 1998). 
The context of this model is  the quasi-spherical accretion model proposed 
by Collin-Souffrin et al. (1996). 
The line profile and spectral continuum are computed by coupling 
 a transfer code (TITAN) adapted to Thomson-thick media   
and a Compton scattering Monte-Carlo code (NOAR). A detailed description 
of these codes is found in Dumont et al. (1999).
Iterations are done between both codes until convergence between the 
ionisation state and the spectral energy distribution inside the shell. 
The Compton code provides the transfer code with the local Compton 
gains and losses and with an amplified 'effective' primary 
(Abrassart \& Dumont 1999). Reciprocally the temperature and 
ionisation structure are provided by the transfer code. 

\begin{figure}[h]
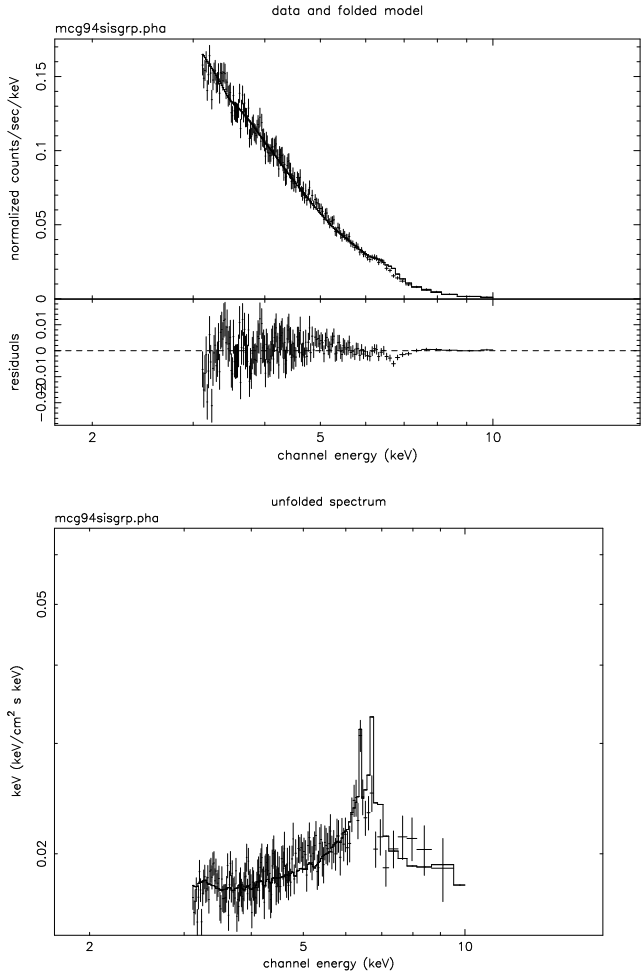

\includegraphics[width=6.5cm,angle=-90]{mmfig3a.ps}
\includegraphics[width=6.5cm,angle=-90]{mmfig3b.ps}
\caption{{\it Up}: 1994 ASCA SIS data of MCG-6-30-15 fitted with the model 
described in the text with residuals plotted below. {\it Bottom}: 
corresponding unfolded spectrum.}
\end{figure}

We have applied this model to the intensively observed source MCG-6-30-15.
Abrassart (1998) showed that
in order to ensure a Compton hump strong enough, as shown by RXTE data, 
cold material is requested
while the red wing requires the presence of ionised material. A high 
covering factor ($\sim\,0.9$) is  needed. 
This model also applies to the ASCA data at higher spectral resolution than 
the RXTE ones. A satisfactory fit is obtained for the 
1994 SIS observations, using an incident spectrum described by the average
AGN distribution derived by Laor et al. (1997), a high covering factor of 
0.9, and clouds 
of different ionisation states (30\% highly ionized 
($\xi$=30000\,erg.cm.s$^{-1}$) and 
70\% near neutral ($\xi$=300)) (see fig. 3).  The red wing of the
iron line is well reproduced while the model predicts an excess of 
highly ionised Fe emission at 6.9 keV which could be suppressed with an even 
more highly ionised medium.  The presence of a two-phase medium might be the
result of a disk disruption leading to a high density contrast.    


\section{Perspectives}
X-ray emission in AGN is a powerful  tool to investigate  
the  very close environnement of the putative black hole. 
As shown above, while present data cannot univocally constrain the 
origin of the 
Fe K$\alpha$ line and the physical parameters of the warm absorber
medium, the present/near-future  generation of X-ray satellites 
(Chandra, XMM, Astro-E) will offer  
an unprecedented high spectral resolution and a much higher 
sensitivity. This will provide an unique opportunity to give access
to the soft X-ray emission lines expected to arise from the WA. 
However this detailed observational information should be supplemented 
with theoretical improvement (precise computation of X-ray atomic data,
computation of realistic disc structure, non-stationary hydrodynamics models 
including possible shocks).

Variability studies on short timescales of the intensity 
and of the profile of the  Fe K$\alpha$ line are crucial to  unambiguously 
confirm the
existence of an accretion disc, to determine the spin of the black hole
and to derive the ionisation structure of the innermost regions. But the 
richness of the data will place greater demands on computations
(such as  which still suffers
from the lack of a consistent model accounting for the complete UV, soft and
hard X-ray spectral distributions.


\end{document}